\documentclass[review]{elsarticle}

\usepackage{lineno,hyperref}
\usepackage{slashed}
\usepackage{amsmath}
\modulolinenumbers[5]

\journal{Physics Letters B}

\begin{document}

\begin{frontmatter}

\title{Massive photon and fermion mixing as manifestations of an extra discrete dimension in bilayer systems}


\author[mymainaddress,mysecondaryaddress]{Nguyen Ai Viet \corref{mycorrespondingauthor}}
\ead{support@elsevier.com}
\cortext[mycorrespondingauthor]{Corresponding author}
\author[mythirdaddress]{Pham Tien Du}
\ead{www.elsevier.com}

\address[mymainaddress]{Faculty of Information Technology, Dai Nam University, Hanoi, Vietnam}
\address[mysecondaryaddress]{ Information Technology Institute, Vietnam National University, Hanoi, Vietnam}
\address[mythirdaddress]{Physics Department, Thuy Loi University, Hanoi, Vietnam}

\begin{abstract}
Discrete dimension is introduced via an extended Dirac operator to include the interlayer interactions in a geometric framework of generic $d+1$-dimensional bilayer systems.
The photon and its Kaluza-Klein partners in this extended space-time will involve a pair of vector fields together with a scalar field, which describes a pair of local jumping between the layers. The quartic potential of the scalar field triggers an abelian Higgs mechanism, which gives a mass to one vector field, which describes a short-range interaction. The discrete derivative along the extra dimension generates a non-diagonal mass matrix, whose eigenstates mix the fermions on two layers with a mixing angle. The couplings of the Kaluza-Klein siblings of photon including massless, massive, and scalar ones with the mass eigenstates of fermions will also depend on the mixing angle. Thus interlayer interactions in bilayer systems are manifested elegantly by these new features via the extra discrete dimension. In specific cases, the short-range force, weak and strong coupling regimes of the fermions with different masses can lead to new physical consequences to be discovered in different realistic $2+1$ and $3+1$-dimensional bilayer systems.    
\end{abstract}


\end{frontmatter}

\linenumbers

\section{Dirac operator and discrete dimensions in bilayer systems}
 
Bilayer systems are common in both particle and condensed matter physics. From the traditional geometric point of view, the space-time in those systems can be perceived just as a direct product of a space-time manifold and a set of two discrete points, which does not change the number of dimensions. However, based on a more advanced geometric concept, one can consider this discrete set as a new dimension on an equal footing and in parallelism with the traditional continuous ones when the Dirac operator is extended with discrete derivatives \cite{Viet1994,Viet1996a,Viet1996b,Viet2003,VDHW2017,Alis2001}.

In particle physics, Connes and Lott \cite{CoLo1989} has proposed that the chiral quark-leptons must be arranged in two different sheets of space-time according to their chiralities. It is the Dirac operator, which identifies the discrete set of two points as a new dimension. In Connes-Lott's model, the usual Dirac operator  $\slashed \partial = \gamma^\mu \slashed \partial_\mu, \mu =0,1,2,3$ is extended by a discrete derivative to
\begin{eqnarray} \label{DiracOperatorChiral}
	\slashed D &=& \Gamma^M D_M = \begin{bmatrix}
		\slashed \partial & i\gamma^5 m \\
		- i \gamma^5 m & \slashed \partial
	\end{bmatrix}, M = 0,1,2,3,4, \nonumber \\
     \Gamma^\mu &=& \gamma^\mu \otimes {\bf 1}_2,~~ \Gamma^4 = \gamma^5 \otimes \sigma_2,
\end{eqnarray}
where $\gamma^\mu, \gamma^5$ are the usual Dirac matrices, ${\bf 1}_2$ and $\sigma_a, a=1,2,3$ are respective $2 \times 2$ unit and Pauli ones.

The set of $\Gamma^M$ forms a basis of $4+1$-d Dirac matrices because
\begin{equation} \label{Clifford}
	\{\Gamma^M, \Gamma^N\} = 2 diag (-1,1,1,1,1)
\end{equation}

The Dirac operator acts on the following reprensentation of chiral spinors
\begin{equation}
	\Psi(\hat x) = \begin{bmatrix}
		\psi_L(x) \\
		\psi_R(x)
		\end{bmatrix}. 
\end{equation}

So, with the chiral index $I=L,R$ as a layer one, the Standard Model can be considered as a $3+1$-d bilayer system. 

The Dirac Lagrangian defined as follows
\begin{equation}
	{\cal L}_f = Tr {\bar \Psi}(\hat x) ( i \slashed D \Psi(\hat x) = i\bar \psi(x) i\slashed \partial + m) \psi(x), 
\end{equation}
where $\psi(x) = \psi_L(x) + \psi_R(x)$ represents a Dirac fermion of mass $m$.  

In Connes-Lott's model, the Higgs field arises naturally as a Kaluza-Klein partner of the gauge fields. In some sense, the left-handed chiral quark-leptons are Kaluza-Klein partners of the right-handed ones. Another example is the Randall-Sundrum model \cite{RaSu1999, RaSu1999a}, two specific warped membranes, which accommodate the elementary particles, are postulated, giving an elegant solution to the naturalness problem. The world in the Randall-Sundrum model can also be considered as another type of bilayer system embedded in five-dimensional space-time.
The concept of discrete dimensions has been also considered in various contexts \cite{ADD1998,ASM2004,Deffayet2004,Deffayet2005}.

In $2+1$-dimensional space-time, the description of bilayer Quantum Hall systems\cite{Viet1996c} in condensed matter physics has been one of the first attempts to use the discrete dimension to include the interlayer interactions. In this formulation, the Chern-Simon term governing the dynamics of these systems is naturally extended to include a complex scalar field representing the tunneling of electrons between the two layers. From the physical point of view, the real 2+1-dimensional systems are not perfect ones. The $3+1$-dimensional effects must be included gradually and manually to explain the observations. In bilayer systems, these effects are more important. By introducing an extra dimension one takes the interlayer interaction into account. Thus this geometric formulation provides an elegant framework to derive more complete physical dynamics in a given bilayer system.

\section{$d+1$-dimensional bilayer system}
In this article, although all results can be extended to the case of an arbitrary $d$, we are interested mainly in the realistic cases of $d=2$ and $d=3$.
\paragraph{Discrete dimension encoded in the Dirac operator}
The Dirac operator given in Eq.(\ref{DiracOperatorChiral}) is not the only possible one. Especially, in the $2+1$-dimensional space-time, the chiral matrix does not exist. The new Dirac operator with a new discrete dimension, which is not chiral and can be defined in any $d+1$-dimensional space-time will bring about new interesting features, which are originated from the new interlayer interactions.  

Let us consider a bilayer system with two different fermion fields $\psi_i(x)$ of mass $m_i$ defined on each layer indexed by $i=1,2$. $x$ now can be coordinates of a $2+1$- or $3+1$-dimensional space-time. The original Lagrangian of these two spinor fields is given as follows
\begin{equation}
	{\cal L}_{0} = \sum_i \bar \psi(x) ( i\slashed \partial + m_i ) \psi(x),
\end{equation}  
where $m_i$ is the mass of $\psi_i(x)$. 

Let us introduce a new extra dimension of the index $d$ via the Dirac operator as follows
\begin{eqnarray} \label{DiracOperatorDark}
	\hat {\slashed D} &=& \hat \Gamma^M D_M= \begin{bmatrix}
		\slashed \partial &  {\bf 1} m \\
		- {\bf 1} m & \slashed \partial
	\end{bmatrix},~~ M=\mu, d,\nonumber \\
\hat \Gamma^\mu &=& \gamma^\mu \otimes \sigma_3,~~\hat \Gamma^d = {\bf 1} \otimes {\bf 1} \otimes \sigma_2,~~ \mu =0,1,..,d-1, \nonumber \\
D_\mu &=& \partial_\mu \sigma_3,~~ D_5= i m \sigma_3,   
\end{eqnarray}
where $\gamma^\mu$ are the Dirac matrices of the d-dimensional space-time.

It is straightforward to verify that the matrices $\hat \Gamma^M$ also satisfy the Clifford algebra
\begin{equation}
	\{\hat \Gamma^M, \hat \Gamma^N\} = 2 diag(-1,1,1,...,1),
\end{equation}
to form a Dirac matrix basis of the $d+1$-dimensional space-time.

The Dirac operator $\hat {\slashed D}$ acts on the bilayer spinor represented by the following column array
\begin{equation} \label{spinor}
		\Psi(\hat x) = \begin{bmatrix}
		\psi_1(x) \\
		\psi_2(x)
	\end{bmatrix}, 
\end{equation}
where $\psi_i(x), i=1,2$ are Dirac spinors. $\hat x$ represents the generalized coordinates of the $d+1$-dimensional space-time including the discrete one, while $x$ represents the coordinates of the d-dimensional space-time.

\paragraph{Photon's Kaluza-Klein partners}

The photon's vector field $A_\mu(x)$ in a $d+1$-d bilayer system, can be extended to the following matrix acting on the generalized spinor in Eq.(\ref{spinor})
\begin{eqnarray}
	\slashed B(\hat x) &=& \hat \Gamma^M B_M(\hat x) = \begin{bmatrix}
		\slashed b_1 (x) & i \phi (x) \\
		-i \phi(x) & \slashed b_2(x)
	\end{bmatrix}, \nonumber \\
	B_\mu(\hat x) &=& \begin{bmatrix}
		b_{1\mu}(x) & 0 \\
		0 & - b_{2\mu} (x)
	\end{bmatrix},~~ B_d(\hat x) = \phi(x)
\end{eqnarray}
where $b_{i\mu}(x)$ are the electromagnetic vector fields of two layers. The scalar field $\phi(x)$ describes interlayer interactions.
We can also represent the vector fields $b_{i\mu}(x) $ as follows
\begin{eqnarray}
	b_{1\mu}(x) &=& g A_\mu(x), \nonumber \\
	b_{2\mu}(x) &=& g A_\mu + g_X X_\mu(x).
\end{eqnarray}

The field strength of the bilayer system is 
\begin{equation} \label{GaugeB}
	{\cal B} = \hat \Gamma^M \wedge \hat \Gamma^N {\cal B}_{MN}(\hat x) =\hat {\slashed D} \wedge \slashed B(\hat x) + \slashed B(\hat x) \wedge \slashed B(\hat x).
\end{equation}
The wedge product is defined as a direct generalization of the usual one in order to generate the Higgs quartic potential
\begin{eqnarray}
	{\hat \Gamma}^\mu \wedge {\hat \Gamma}^\nu &=& -{\hat \Gamma}^\nu \wedge \hat \Gamma^\mu,~~{\hat \Gamma}^\mu \wedge {\hat \Gamma}^d = {\hat \Gamma}^d \wedge {\hat \Gamma}^\mu \nonumber \\
	{\hat \Gamma}^d \wedge {\hat \Gamma}^d &=& 1.
\end{eqnarray}

Therefore, the field strength's components are calculated as follows
\begin{subequations}
	\begin{alignat}{2}
		&{\cal B}_{\mu \nu} = {1 \over 2} (\partial_\mu B_\nu(x) - \partial_\nu B_\mu(x) + [B_\mu(x), B_\nu(x)])  \\
		&{\cal B}_{\mu d} = {1 \over 2}  (\partial_\mu + g' X_\mu(x) \sigma_3)  (\varphi(x) + m)  \\
		&{\cal B}_{dd} = ~~2m \varphi(x) \sigma_3 + \varphi^2(x) {\bf 1}_2. 
	\end{alignat}
\end{subequations}

The full Lagrangian of the generalized vector field is given as follows
\begin{eqnarray}
	{\cal L}_g &=& - {1 \over 2 g^2} Tr({1 \over 4} {\cal B}^{\mu \nu} {\cal B}_{\mu \nu} + {1 \over 2} G^{\mu \nu}{\cal B^*}_{\nu d} B_{\mu d} \nonumber \\
	&& + {\cal B^*}_{dd} {\cal B}_{dd} ). 
\end{eqnarray}

The field strength of the vector fields and the physical scalar field $H(x)$  are defined as follows
\begin{eqnarray}
	&&F_{\mu \nu} = \partial_\mu A_\nu(x) - \partial_\nu A_\mu(x),~X_{\mu \nu} = \partial_\mu X_\nu(x) - \partial_\nu X_\mu(x) \nonumber \\
	&&\varphi(x) = \sqrt{2} g H(x) - m .
\end{eqnarray}
So, we will conveniently call the quanta of the physical fields $X_\mu(x)$ and $H(x)$ respectively as X- and H-photons, since they are Kaluza-Klein partners of the photon.

The Lagrangian for the gauge sector now is
\begin{eqnarray}
	&&{\cal L}_g= - { 1 \over 2g^2  } Tr (B^{\mu \nu} B_{\mu \nu} + 2 B_{\mu 5} B_{\nu 5} 
	+  B^2_{55}), \nonumber \\
	&&=  ( - {1\over 4} F^{\mu \nu} F_{\mu \nu} - {g_X^2 \over 8 g^2} X^{\mu \nu} X_{\mu \nu}+ {1 \over 2 } \partial^\mu H(x) \partial_\mu H(x) \nonumber \\
	& &+  g_X^2 X^2_\mu(x) H^2(x) - 2 g^2 (H^2(x) - {m^2 \over 2 g^2 })^2). \label{Gauge}
\end{eqnarray}
In order to have the correct kinetic term for the X-photon the coupling constant $g_X$ must be expressed in term of the electromagnetic one as follows
\begin{equation} \label{gandg}
	g_X = \sqrt{2} g. 
\end{equation}

Due to the quartic potential in Eq.(\ref{Gauge}), the H-photon field $H(x)$ has a vacuum expectation value (VEV) $v= m/\sqrt{2} g$, giving the X-photon a mass $ m_X = m$. The H-photon has a mass $ \sqrt{2} m$. This is an Abelian Higgs mechanism where the X-photon receives a mass from H-photon's VEV. The renormalizability is kept intact.
The normal photon field $A_\mu(x)$ remains massless. 

\paragraph{Non-diagonal mass matrix}

The Dirac Lagrangian of a pair of fermion extends the usual one $\bar \psi (x) (i \slashed \partial + m_0) \psi(x)$ as follows
\begin{eqnarray}\label{FermionLagrangian}
	{\cal L}_D &=& Tr (\bar \Psi (\hat x) (i \hat {\slashed D} +  M_0) \Psi(\hat x)) \nonumber \\
	&=& \sum_i \bar \psi_i(x) i \slashed \partial \psi_i(x) + \sum_{i,j} \bar \psi_i {\cal M}_{ij} \psi_j(x)  , 
\end{eqnarray} 
where
\begin{equation} \label{ModMassMatrix}
	M_0 = \begin{bmatrix}
		m_1 & 0 \\
		0 & m_2
	\end{bmatrix},~~
{\cal M} = \begin{bmatrix}
	m_1 & i m  \\
	- i m  & m_2
\end{bmatrix}.
\end{equation}
The discrete derivative has contributed to the off-diagonal elements of the hermitian mass matrix ${\cal M}$. 

\paragraph{Fermion mixing and mass eigenstates}

The presence of non-diagonal elements in the mass matrix ${\cal M}$ is a consequence of the discrete dimension. Thus, the spinors $\psi_i(x)$  are not mass eigenstates due to the interlayer interactions. We can diagonalize it to have the mass eigenstates by the following transformation

\begin{equation} \label{UTrans}
	\Psi'(\hat x) = U \Psi(\hat x) = \begin{bmatrix}
		\psi(x) \\
		\psi_X(x)
	\end{bmatrix},~~ U = \begin{bmatrix}
		\cos \theta & i\sin \theta \\
		-i \sin \theta & -\cos \theta 
	\end{bmatrix},
\end{equation}
where $\psi(x)$ and $\psi_X(x)$ are two mass eigenstates of the fermion's Kaluza-Klein partners.The Dirac operator in Eq.(\ref{DiracOperatorDark}) has been chosen so that the mass matrix ${\cal M}$ is hermitian.
 
The unitary transformation $U$ will mix the components $\psi_i(x)$ of the generalized spinor  $\Psi(x)$ in Eq.(\ref{spinor}) as follows
\begin{equation} \label{MixingTrans}
	\Psi(\hat x) \rightarrow U\Psi(\hat x),~~ \bar \Psi(\hat x) \rightarrow \bar \Psi(\hat x) U^{\dagger}.
\end{equation}

The kinetic term of the free fermion's Lagrangian in Eq.(\ref{FermionLagrangian}) is invariant with the transformation $U$. The mass term is transformed to the following diagonalized matrix
\begin{equation} \label{MassMatrix}
	{\cal M}' = U {\cal M} U^{\dagger} = \begin{bmatrix}
		\mu & 0 \\
		0 & \mu_D
	\end{bmatrix},
\end{equation}  
where $\mu$ and $\mu_D$ are two real mass eigenvalues.

Inversely, the original mass matrix can be rewritten as follows
	\begin{equation} \label{MassTransform}
		M = U^\dagger M' U = 
		\begin{bmatrix}
			\mu + (\mu_D - \mu) \sin^2 \theta & {i \over 2} (\mu_D-\mu) \sin 2 \theta   \\
			{-i \over 2} (\mu_D-\mu) \sin 2 \theta  & \mu + (\mu_D-\mu) \cos^2 \theta 
		\end{bmatrix}. 
	\end{equation}

Comparing Eqs.(\ref{MassTransform}) and (\ref{ModMassMatrix}), we obtain the following mass splitting formula for the fermion Kaluza-Klein pair's mass eigenstates
\begin{equation} \label{MassSplit}	
	\delta \mu = \mu_D- \mu = {2 m \over \sin 2\theta}.
\end{equation}

So we can express the mass matrix ${\cal M}$ in new parameters $\mu, m$ and $\theta$ as follows
\begin{eqnarray} \label{MassNew}
	M &=& 
	\begin{bmatrix}
		\mu + m \tan \theta   & i m   \\
		-im  & \mu + m \cot \theta
	\end{bmatrix}.
\end{eqnarray}

The mass parameters $m_i$ in Eq.(\ref{MassMatrix}) are represented as follows
\begin{equation}
	m_1= \mu +  m \tan \theta , ~~ m_2 = \mu +  m \cot \theta .
\end{equation}

\paragraph{Couplings of fermion's mass eigenstates to the photon's Kaluza-Klein partners}
The interaction Lagrangian of fermions with the  photon's Kaluza-Klein partners is the following extension of the usual $\bar \psi(x) \slashed A(x) \psi(x)$ 
\begin{equation}
	{\cal L}_{f-g} = Tr \bar \Psi(\hat x) \slashed B (\hat x) \Psi(\hat x) = Tr \bar \Psi(\hat x) U^\dagger ( U \slashed B (\hat x) U^\dagger) U \Psi(\hat x).
\end{equation}

So, the couplings of the fermion mass eigenstates to the photon's Kaluza-Klein partners can be calculated by using the  transformation (\ref{UTrans}) as follows 

\begin{subequations}
	\begin{alignat}{2}
		&{\cal L}_A =  g (\bar \psi(x) \slashed {A} \psi (x) + \bar \psi_X(x) \slashed {A} \psi_X(x)), \label{LA}\\
		&{\cal L}_X =
		\sqrt{2} g (\sin^2 \theta \bar \psi(x) \slashed{X}  \psi(x)
		+\cos^2 \theta \bar \psi_X(x) \slashed{X} \psi_X(x) \nonumber \\
		&+ \frac{1}{{2}} \sin 2\theta \bar \psi_X(x)  \slashed{X} \psi(x) + \frac{1}{{2}}  \sin 2\theta \bar \psi(x) \slashed{X} \psi_X(x)), \label{LX}
		\\
		&{\cal L}_H =
		i\sqrt{2}g \sin 2 \theta (H(x) - {m \over \sqrt{2} g})({\bar \psi}(x) \psi(x) - {\bar \psi}_X(x) \psi_X(x)), \nonumber \\
		& 
		+ i \sqrt{2}g \cos 2\theta (H(x) - {m \over \sqrt{2} g}) ({\bar \psi}(x) \psi_X(x) - \bar \psi_X(x) \psi(x))). 
		\label{LH}
	\end{alignat}
\end{subequations}

The Lagrangian ${\cal L}_A$ in Eq.(\ref{LA}) represents the usual electromagnetic interaction of the fermions $\psi(x)$ and $\psi_X(x)$ with the same coupling constant, meaning that the fermion Kaluza-Klein partners have the same electric charge. The Lagrangian ${\cal L}_X$ represents the interaction of $X$-photon with fermions. The X-coupling of $\psi$ with X is $\sqrt{2} g \sin^2\theta$, while the one of $\psi_X$ is $\sqrt{2} g \cos^2\theta$. Depending on the mixing angle, the X-coupling of $\psi$ is small when the one of $\psi_X$ is large and vice versa. There are portals between the fermion Kaluza-Klein partners $\psi$ and $\psi_X$ via the decays involving the X- and H-photons. 
  
\section{Summary and discussions}

In this paper, we have presented a geometric framework based on an extension of the $d+1$-dimensional space-time with a set of two points. In this approach, the Dirac operator is extended by a discrete derivative, leading to a notion of new dimension, which hopefully provide a more complete physical dynamics of the interlayer interactions. This description leads to three new features. 

The first one is that the new Dirac operator generates a non-diagonal mass matrix. Therefore, the mass eigeinstates, which mix the fermions on two layers, will have different masses, split by the mixing angle and a mass parameter, which is inverse of the distance between two layers. 

The second feature is that the extended photon field consists of a set of three Kaluza-Klein partner, including the photon fields on two layers and a scalar field, which might describe the interlayer jumpings. Since when a fermion jumps from one layer to the other one, it leaves a hole in the first one, which triggers a fermion on the second layer jumps back to the first one. As a result, two jumpings will form a local order parameter, which represents the scalar Kaluza-Klein partner of the photon fields. The model construction procedure inspired by noncommutative geometry (NCG) brings about a quartic potential of this scalar field, which trigger an abelian Higgs mechanism, giving mass to the difference between photon fields on two layers. As a consequence, we have two vector fields, one is massless, representing a long range force, the other is massive, representing a short range one. 

The third feature of this description is that the couplings of the massive and scalar photons to the fermions mass eigenstates are governed by the mixing angle. When this angle becomes small, the couplings of the X-photon will be small to one fermion and will be larger to the other one.  

The above features might lead to rich physical consequences in the realistic $2+1$- and $3+1$-dimensional bilayer systems. Let us illustrate with a few simple first ideas. In the realistic $2+1$-dimensional condensed matter bilayer systems, since the typical interlayer distance can be achieved as the atomic size, the mass of X- and H-photons can be up to $10-100~eV$. If the fermion mass splitting is comparable to the electron mass, for example, being a few hundred $keV$, the mixing angle $\theta \sim 10^{-4} - 10^{-3}$. Then the third feature might be explored.

In the $3+1$-dimensional nuclear and high energy physics bilayer systems, the detection of the massive boson and fermions will shed light on the existence and size of the extra dimension. For instance, the hypothetical $X17$ boson of $17~MeV$ was reportedly observed in the IPC experiment at ATOMKI \cite{Atomki2015, Atomki2018, Atomki2019} can be identified as H- or X-photon of this picture, indicating the discrete dimension of size $11.8~ fm$. The second idea is to utilize a discrete dimension to provide a possible explanation for the differences in the coupling constants of different interactions. In this picture, the electroweak coupling constant of the Kaluza-Klein partners will be strong, while the strong one will be weak if these particles will be found in the $TeV$ or higher energy range. The works along these directions are in progress.
 
\section*{Acknowledgement}
The research is funded by Vietnam National Foundation for Science and Technology Development(NAFOSTED) under grant No 103.01-2017.319.

\bibliography{FermionMixing}

\begin{thebibliography}{10}
\expandafter\ifx\csname url\endcsname\relax
  \def\url#1{\texttt{#1}}\fi
\expandafter\ifx\csname urlprefix\endcsname\relax\def\urlprefix{URL }\fi
\expandafter\ifx\csname href\endcsname\relax
  \def\href#1#2{#2} \def\path#1{#1}\fi

\bibitem{Viet1994}
G.~Landi, N.~A. Viet, K.~C. Wali, {Gravity and electromagnetism in
  noncommutative geometry}, Phys. Lett. B326 (1994) 45--50.
\newblock \href {http://arxiv.org/abs/hep-th/9402046}
  {\path{arXiv:hep-th/9402046}}, \href
  {https://doi.org/10.1016/0370-2693(94)91190-8}
  {\path{doi:10.1016/0370-2693(94)91190-8}}.

\bibitem{Viet1996a}
N.~A. Viet, K.~C. Wali, {Noncommutative geometry and a discretized version of
  Kaluza-Klein theory with a finite field content}, Int. J. Mod. Phys. A11
  (1996) 533--552.
\newblock \href {http://arxiv.org/abs/hep-th/9412220}
  {\path{arXiv:hep-th/9412220}}, \href
  {https://doi.org/10.1142/S0217751X96000249}
  {\path{doi:10.1142/S0217751X96000249}}.

\bibitem{Viet1996b}
N.~A. Viet, K.~C. Wali, {A Discretized version of Kaluza-Klein theory with
  torsion and massive fields}, Int. J. Mod. Phys. A11 (1996) 2403--2418.
\newblock \href {http://arxiv.org/abs/hep-th/9508032}
  {\path{arXiv:hep-th/9508032}}, \href
  {https://doi.org/10.1142/S0217751X96001206}
  {\path{doi:10.1142/S0217751X96001206}}.

\bibitem{Viet2003}
N.~A. Viet, K.~C. Wali, {Chiral spinors and gauge fields in noncommutative
  curved space-time}, Phys. Rev. D67 (2003) 124029.
\newblock \href {http://arxiv.org/abs/hep-th/0212062}
  {\path{arXiv:hep-th/0212062}}, \href
  {https://doi.org/10.1103/PhysRevD.67.124029}
  {\path{doi:10.1103/PhysRevD.67.124029}}.

\bibitem{VDHW2017}
N.~A. Viet, N.~V. Dat, N.~S. Han, K.~C. Wali, {Einstein-Yang-Mills-Dirac
  systems from the discretized Kaluza-Klein theory}, Phys. Rev. D95~(3) (2017)
  035030.
\newblock \href {http://arxiv.org/abs/1611.01738} {\path{arXiv:1611.01738}},
  \href {https://doi.org/10.1103/PhysRevD.95.035030}
  {\path{doi:10.1103/PhysRevD.95.035030}}.

\bibitem{Alis2001}
M.~Alishahiha, {(De)constructing dimensions and noncommutative geometry}, Phys.
  Lett. B517 (2001) 406--414.
\newblock \href {http://arxiv.org/abs/hep-th/0105153}
  {\path{arXiv:hep-th/0105153}}, \href
  {https://doi.org/10.1016/S0370-2693(01)00984-4}
  {\path{doi:10.1016/S0370-2693(01)00984-4}}.

\bibitem{CoLo1989}
A.~Connes, J.~Lott, {Particle Model and Noncommutative Geometry}, Nuclear
  Physics (Proc.Suppl.) 18B (1990) 29--47.

\bibitem{RaSu1999}
L.~Randall, R.~Sundrum, {A Large mass hierarchy from a small extra dimension},
  Phys. Rev. Lett. 83 (1999) 3370--3373.
\newblock \href {http://arxiv.org/abs/hep-ph/9905221}
  {\path{arXiv:hep-ph/9905221}}, \href
  {https://doi.org/10.1103/PhysRevLett.83.3370}
  {\path{doi:10.1103/PhysRevLett.83.3370}}.

\bibitem{RaSu1999a}
L.~Randall, R.~Sundrum, {An Alternative to Compactification}, Phys. Rev. Lett.
  83 (1999) 4690--4693.
\newblock \href {http://arxiv.org/abs/hep-th/9906064}
  {\path{arXiv:hep-th/9906064}}, \href
  {https://doi.org/10.1103/PhysRevLett.83.4690}
  {\path{doi:10.1103/PhysRevLett.83.4690}}.

\bibitem{ADD1998}
N.~Arkani-Hamed, S.~Dimopoulos, G.~R. Dvali, {The Hierarchy problem and new
  dimensions at a millimeter}, Phys. Lett. B429 (1998) 263--272.
\newblock \href {http://arxiv.org/abs/hep-ph/9803315}
  {\path{arXiv:hep-ph/9803315}}, \href
  {https://doi.org/10.1016/S0370-2693(98)00466-3}
  {\path{doi:10.1016/S0370-2693(98)00466-3}}.

\bibitem{ASM2004}
N.~Arkani-Hamed, M.~D. Schwartz, {Discrete gravitational dimensions}, Phys.
  Rev. D69 (2004) 104001.
\newblock \href {http://arxiv.org/abs/hep-th/0302110}
  {\path{arXiv:hep-th/0302110}}, \href
  {https://doi.org/10.1103/PhysRevD.69.104001}
  {\path{doi:10.1103/PhysRevD.69.104001}}.

\bibitem{Deffayet2004}
C.~Deffayet, J.~Mourad, {Multigravity from a discrete extra dimension}, Phys.
  Lett. B589 (2004) 48--58.
\newblock \href {http://arxiv.org/abs/hep-th/0311124}
  {\path{arXiv:hep-th/0311124}}, \href
  {https://doi.org/10.1016/j.physletb.2004.03.053}
  {\path{doi:10.1016/j.physletb.2004.03.053}}.

\bibitem{Deffayet2005}
C.~Deffayet, J.~Mourad, {Deconstruction of gravity}, Int. J. Theor. Phys. 44
  (2005) 1743--1752.
\newblock \href {https://doi.org/10.1007/s10773-005-8892-0}
  {\path{doi:10.1007/s10773-005-8892-0}}.

\bibitem{Viet1996c}
V.~John, N.~A. Viet, K.~C. Wali, {Chern-Simons terms in noncommutative geometry
  and its application to bilayer quantum Hall systems}, Phys. Lett. B371 (1996)
  252--260.
\newblock \href {http://arxiv.org/abs/hep-th/9503091}
  {\path{arXiv:hep-th/9503091}}, \href
  {https://doi.org/10.1016/0370-2693(95)01608-2}
  {\path{doi:10.1016/0370-2693(95)01608-2}}.

\bibitem{Atomki2015}
A.~J. Krasznahorkay, M.~Csatlós, L.~Csige, et~al., {Observation of Anomalous
  Internal Pair Creation in Be8 : A Possible Indication of a Light, Neutral
  Boson}, Phys. Rev. Lett. 116~(4) (2016) 042501.
\newblock \href {http://arxiv.org/abs/1504.01527} {\path{arXiv:1504.01527}},
  \href {https://doi.org/10.1103/PhysRevLett.116.042501}
  {\path{doi:10.1103/PhysRevLett.116.042501}}.

\bibitem{Atomki2018}
A.~J. Krasznahorkay, M.~Csatlós, L.~Csige, et~al., {New results on the $^8$Be
  anomaly}, J. Phys. Conf. Ser. 1056~(1) (2018) 012028.
\newblock \href {https://doi.org/10.1088/1742-6596/1056/1/012028}
  {\path{doi:10.1088/1742-6596/1056/1/012028}}.

\bibitem{Atomki2019}
A.~J. Krasznahorkay, M.~Csatlós, L.~Csige, et~al., {On the $X$(17)
  Light-particle Candidate Observed in Nuclear Transitions}, Acta Phys. Polon.
  B50~(3) (2019) 675.
\newblock \href {https://doi.org/10.5506/APhysPolB.50.675}
  {\path{doi:10.5506/APhysPolB.50.675}}.

\end{thebibliography}

\end{document}